%% file: iccd.tex
\documentclass[conference]{IEEEtran}
\IEEEoverridecommandlockouts
\usepackage{cite}

\ifCLASSINFOpdf
   \usepackage[pdftex]{graphicx}
\else
\fi
\usepackage{algorithm}
\usepackage[noend]{algpseudocode}
\usepackage{color}
\usepackage{colortbl}
\usepackage{multirow}
\usepackage{subcaption}
\usepackage{microtype}
\usepackage{booktabs}
\usepackage[inline]{enumitem}
\usepackage{xcolor}
\usepackage{hyperref}
\hypersetup{
	breaklinks=true,
	hidelinks=true
}
\usepackage{eso-pic}

\makeatletter
\def\ps@IEEEtitlepagestyle{%
	\def\@oddfoot{\mycopyrightnotice}%
	\def\@oddhead{\hbox{}\@IEEEheaderstyle\leftmark\hfil\thepage}\relax
	\def\@evenhead{\@IEEEheaderstyle\thepage\hfil\leftmark\hbox{}}\relax
	\def\@evenfoot{}%
}
\def\mycopyrightnotice{%
	\begin{minipage}{\textwidth}
		\scriptsize
		\copyright~2024 IEEE. DOI: 10.1109/ICCD63220.2024.00108. Personal use of this material is permitted. 
		Permission from IEEE must be obtained for all other uses,
		in any current or future media, including reprinting/republishing this material
		for advertising or promotional purposes, creating new collective works,
		for resale or redistribution to servers or lists, or reuse of any copyrighted
		component of this work in other works by sending a request to pubs-permissions@ieee.org.
	\end{minipage}
}
\makeatother

\begin{document}
%
\title{A Prototype-Based Framework to Design Scalable Heterogeneous SoCs with Fine-Grained DFS
\thanks{This work was partly funded by the ISOLDE project (Grant No. 101112274), supported by CHIPS Joint Undertaking.}}

\author{\IEEEauthorblockN{Gabriele Montanaro, Andrea Galimberti, Davide Zoni}
\IEEEauthorblockA{\textit{Dipartimento di Elettronica, Informazione e Bioingegneria (DEIB)}\\
\textit{Politecnico di Milano}, Milan, Italy\\
\{gabriele.montanaro, andrea.galimberti, davide.zoni\}@polimi.it}}


%


\maketitle

\begin{abstract}
Frameworks for the agile development of modern system-on-chips are
crucial to dealing with the complexity of designing such architectures.
The open-source Vespa framework for designing large,
FPGA-based, multi-core heterogeneous system-on-chips enables a faster and
more flexible design space exploration of such architectures and
their run-time optimization.
Vespa, built on ESP, introduces
the capabilities to instantiate multiple replicas of
the same accelerator in a single network-on-chip node and to partition
the system-on-chips into frequency islands with independent dynamic frequency scaling actuators,
as well as a dedicated run-time monitoring infrastructure.
Experiments on 4-by-4 tile-based system-on-chips demonstrate
the possibility of effectively exploring a multitude of solutions that
differ in the replication of accelerators, the clock frequencies of the frequency islands,
and the tiles' placement, as well as
monitoring a variety of statistics related to the traffic on the interconnect and
the accelerators' performance at run time.
\end{abstract}


%
\IEEEpeerreviewmaketitle

\section{Introduction}
\label{sec:introduction}
\input{01_introduction.tex}

\section{Methodology}
\label{sec:methodology}
\input{02_methodology.tex}

\section{Experimental evaluation}
\label{sec:expEval}
\input{03_experiments.tex}

\section{Conclusions}
\label{sec:conclusions}
\input{04_conclusions.tex}




\bibliographystyle{IEEEtran}
\bibliography{IEEEabrv,iccd}

%
%
%

\end{document}

%% file: 01_introduction.tex
While general-purpose central processing units~(CPUs) transitioned to multi- and many-core architectures
in the last decades, due to the slowdown in the improvement of
their performance and efficiency with the end of Moore's law and Dennard scaling,
hardware acceleration emerged as the standard solution to support
modern computationally intensive workloads ranging from cryptography~\cite{Galimberti_2022DSD}
to deep-learning~\cite{Chen_2016JSSC} ones.

The increased complexity and development costs of system-on-chips~(SoCs),
that are commonly heterogeneous multi-core processors combining
general-purpose CPU cores and hardware accelerators,
including those obtained through high-level synthesis~(HLS)~\cite{Galimberti_2023ICECS},
have imposed the research and the adoption of novel design methods for their architectural exploration,
system integration, verification, validation, and physical design.
Agile development through field-programmable gate array~(FPGA)-based prototyping has therefore emerged as
the standard paradigm to support the rapid design of
complex heterogeneous platforms, offering
a complementary tool to classic cycle-accurate simulators.
For this reason, a multitude of frameworks to quickly deliver FPGA prototypes has been
introduced during the last decade, fueled by a drop in the cost per look-up table
and by the availability of ever-larger FPGAs.

While cycle-accurate simulators were traditionally used
in the early stages of the architectural exploration, the specialization of
the current computing platforms has hindered their adoption.
They fail indeed to support the accurate simulation of complex computing platforms with
hardware accelerators since the employed generic area and power models cannot
provide accurate estimates, whereas designing area and power simulation models
for each specific accelerator is a lengthy and difficult process.
In addition, simulators' performance is severely curbed
by their single-thread software architecture.

Enabling a quick and comprehensive design space exploration~(DSE) of
such complex architectures, also accounting for their run-time optimization,
is critical, and even the few available open-source frameworks~\cite{mantovani2020agile,amid2020chipyard,openpiton_2020}
lack such support.
Indeed, the ESP~\cite{mantovani2020agile} state-of-the-art framework for the agile development
of complex accelerator-centric multi-core heterogeneous SoCs
notably combines a scalable tile-based architecture and a flexible
system-level design methodology to deliver prototypes on
FPGA targets, but its tiles' limited configurability prevents an in-depth DSE
and it lacks the dynamic frequency scaling~(DFS) actuators and a proper run-time monitoring infrastructure
required to support an effective run-time optimization~\cite{Zoni_2023CSUR}.

\paragraph*{Contributions}
This work introduces the Vespa framework, which extends ESP~\cite{mantovani2020agile}
to enable both the DSE and the run-time optimization of large
multi-core heterogeneous SoCs, providing three main contributions.
\begin{enumerate}
	\item A design-time parameter and related hardware infrastructure enable
	scaling the throughput of third-party hardware accelerators,
	without altering the design of the latter or the network-on-chip~(NoC) interconnect,
	by instantiating multiple replicas of the same accelerator
	within a single computing tile assigned to an NoC node,
	
	\item The on-chip interconnect and the computing elements
	can be configurably partitioned into frequency islands, each of whom is independent from the others
	and is fed a clock signal that is either fixed or generated by a DFS actuator.
	
	\item A run-time monitoring infrastructure enables the collection of various execution statistics,
	exposed through memory-mapped hardware counters, to support the DSE and
	the run-time optimization of the SoC.
\end{enumerate}
The Vespa framework for the design space exploration and run-time optimization of large
multi-core heterogeneous SoCs is publicly available and released
open source\footnote{\url{https://github.com/hardware-fab/vespa}.}
with the goal of fostering further research in the field.

\begin{figure*}[t]
	\centering
	\includegraphics[width=\textwidth]{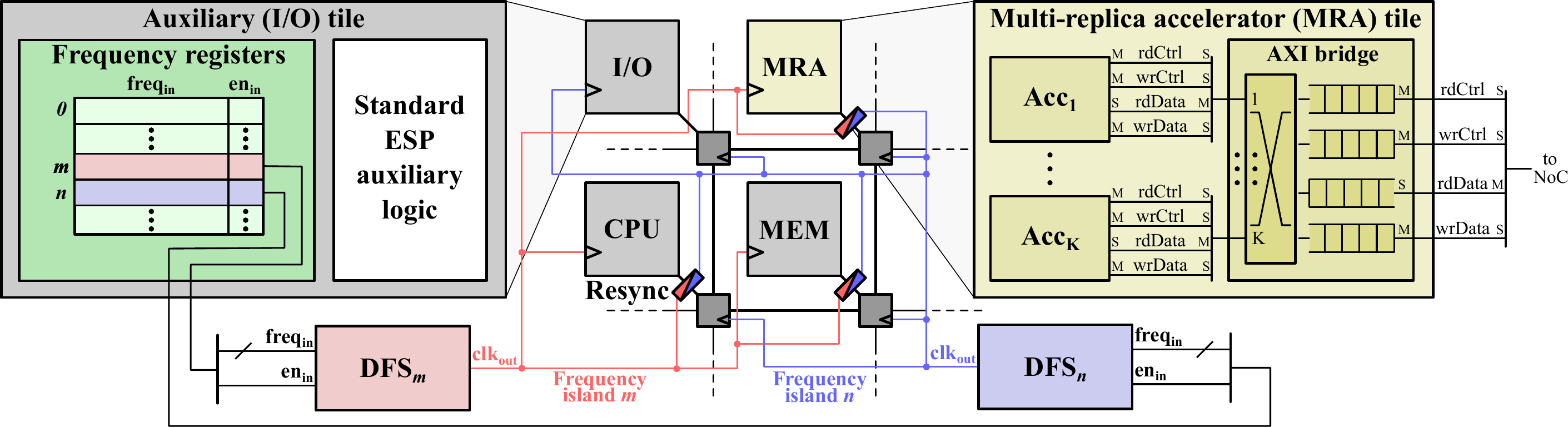}
	\caption{Architecture of a generic Vespa SoC with
		multi-replica accelerator tiles and configurable-DFS frequency islands.
		Example with \emph{CPU}, \emph{MEM}, and \emph{MRA} tiles
		in \emph{Frequency island \textit{m}},
		\emph{I/O} tile and interconnect in \emph{Frequency island \textit{n}},
		resychronizers~(\emph{Resync}) at boundaries of frequency islands.
		Legend: \emph{M} AXI master, \emph{S} slave,
		\emph{K} replication factor of \emph{MRA} tile.}
	\label{fig:soc_our}
\end{figure*}

%% file: 02_methodology.tex
The Vespa framework, depicted in \figurename~\ref{fig:soc_our},
leverages ESP and enables
the design space exploration and run-time optimization of
complex accelerator-based SoCs by means of
flexible multi-replica accelerator tiles,
configurable-DFS frequency islands, and
a dedicated run-time monitoring infrastructure.

\subsection{Multi-replica accelerator-tile architecture}
\label{ssec:meth_accels}
A simple and time-efficient solution to increase the throughput of
an existing hardware accelerator integrated into a computing platform
is to instantiate multiple replicas of the former.
Indeed, instantiating \textit{K} such replicas intuitively provides
a \textit{K\texttimes} increase of the throughput of the baseline accelerator.

Vespa's multi-replica accelerator tiles~(\emph{MRA} in \figurename~\ref{fig:soc_our})
can instantiate multiple replicas~(\emph{Acc\textsubscript{i}}) of
a hardware accelerator within a single computing tile without any changes
either to the accelerator or the size and topology of the NoC interconnect,
and avoiding being forced to occupy multiple computing tiles with
the same accelerator in order to improve its performance. 
The multi-replica accelerator tiles
support any third-party hardware accelerator that exposes an AXI interface.
Their replication factor \emph{K}, which
corresponds to the number of copies of
an accelerator that are instantiated and operate in parallel,
is configurable by the system designer to easily tune the
trade-off between throughput and area.

A baseline accelerator to be integrated into an ESP SoC exposes four AXI4-Stream interfaces
for read control~(\emph{rdCtrl}), write control~(\emph{wrCtrl}),
read data~(\emph{rdData}), and write data~(\emph{wrData}) purposes, respectively.
A multi-replica accelerator tile still exposes the same AXI4-Stream interfaces
towards the NoC interconnect~(\emph{NoC}) to guarantee
full compatibility with the ESP SoC architecture.
The \emph{AXI bridge} component is therefore tasked with multiplexing
the four AXI4-Stream interfaces of each of the \emph{K} accelerator replicas
into four corresponding buffers for the AXI4-Stream interfaces of the tile.

\subsection{Configurable-DFS frequency-island architecture}
\label{ssec:meth_islands}
The Vespa framework allows partitioning the SoC into multiple frequency islands
at design time. Every SoC tile and NoC router is assigned to
a frequency island that can group multiple computing and routing elements.
Each frequency island is fed an independent clock signal that can either
have a fixed frequency or be generated by a dedicated DFS actuator.

The architecture supporting such SoC partitioning can
be split into three main components, namely
a set of registers holding the frequency configuration of each
island~(\emph{Frequency registers} in \figurename~\ref{fig:soc_our}),
a DFS actuator~(\emph{DFS}) for each of the frequency islands, and
resynchronizers~(\emph{Resync}) at the boundaries of the latter.

Notably, the clock signal output by the mixed-mode clock managers~(MMCMs) of an AMD FPGA
remains low during its reconfiguration, thus causing
a clock-gating effect on the computing platform.
The DFS actuator implemented in the Vespa framework avoids such negative effect by
employing two MMCM components. In each DFS actuator, an internal finite-state machine forces
the master MMCM to preserve the current output clock signal while
the slave one is under reconfiguration, after which their roles are swapped. 

\begin{table*}[t]
	\centering
	\caption{FPGA resource utilization and throughput of the baseline accelerator tiles
		and of their 2\texttimes\, and 4\texttimes\, multi-replica instances.
		Legend: \textbf{Accel.} accelerator, \textbf{Thr.} throughput (MB/s),
		\textbf{Incr.} average increment compared to baseline accelerator tile.}
	\scalebox{1}{
		\begin{tabular}{crrrrrrrrrrrrrrr}
			\toprule
			&      \multicolumn{5}{c}{\textbf{Baseline (1\texttimes) accelerator}}      &          \multicolumn{5}{c}{\textbf{2\texttimes-replication instance}}           &          \multicolumn{5}{c}{\textbf{4\texttimes-replication instance}}           \\
			\cmidrule(lr){2-6} \cmidrule(lr){7-11} \cmidrule(lr){12-16}
			\textbf{Accel.} & \textbf{LUT} & \textbf{FF} & \textbf{BRAM} & \textbf{DSP} & \textbf{Thr.} &   \textbf{LUT} &    \textbf{FF} &  \textbf{BRAM} & \textbf{DSP} &  \textbf{Thr.} &   \textbf{LUT} &    \textbf{FF} &  \textbf{BRAM} & \textbf{DSP} &  \textbf{Thr.} \\ \midrule
			adpcm                                      &        10899 &       11720 &            25 &           81 &          1.40 &          16455 &          15158 &             48 &          162 &           2.76 &          27313 &          21780 &             94 &          324 &           5.41 \\
			dfadd                                      &        11268 &       11199 &             2 &            9 &          9.22 &          16988 &          14090 &              2 &           18 &          16.88 &          28599 &          19614 &              2 &           36 &          26.06 \\
			dfmul                                      &         8435 &       10222 &             2 &           25 &          8.70 &          11352 &          12136 &              2 &           50 &          15.07 &          17382 &          15706 &              2 &          100 &          26.06 \\
			dfsin                                      &        16627 &       14997 &             2 &           52 &          0.33 &          27770 &          21686 &              2 &          104 &           0.65 &          50043 &          34804 &              2 &          208 &           1.24 \\
			gsm                                       &         9900 &       11418 &            18 &           62 &          4.61 &          14304 &          14520 &             34 &          124 &           8.90 &          22927 &          20473 &             66 &          248 &          16.67 \\ \midrule
			
			\textbf{Incr.}                                 &              &             &               &              &               & 1.50\texttimes & 1.29\texttimes & 1.36\texttimes &  2.00\texttimes & 1.89\texttimes & 2.49\texttimes & 1.85\texttimes & 2.09\texttimes &  4.00\texttimes & 3.41\texttimes \\ \bottomrule
		\end{tabular}
	}
	\label{tab:chstone_area}
\end{table*}

\subsection{Run-time monitoring infrastructure}
\label{ssec:rt_monitoring}
Vespa's run-time monitoring infrastructure is crucial to collecting statistics from
the computing platform and thus supporting run-time optimization
policies and design space exploration.
For each accelerator, the run-time monitoring infrastructure can selectively
enable the monitoring of up to four different statistics, each corresponding to
a specific counter in the tile: execution time, number of incoming and outgoing packets, and
the round-trip time.
The latter is defined as the time occurring between
a request for data from an accelerator to the main memory through
direct memory access and the ensuing arrival of such data to the accelerator.

The execution time counter automatically resets when the 
accelerator tile starts its computation and stops when the latter is completed,
while the other three counters are reset manually.
All counters instantiated on accelerator tiles expose
memory-mapped registers that can be accessed both via software executing on
CPU cores of the SoC itself and via the USB-to-serial interface that connects
the platform to the host.

%% file: 03_experiments.tex
The experimental campaign targeted a Siemens proFPGA quad~(\textit{mb-4m-r3})
motherboard\footnote{\url{https://www.profpga.com/products/motherboards-overview/profpga-quad}}
equipped with a Virtex-7 2000 FPGA~(\textit{fm-xctv2000t-r2})
module\footnote{\url{https://www.profpga.com/products/fpga-modules-overview/virtex-7-based/profpga-xc7v2000t}},
whose AMD FPGA chip features
1221600 look-up tables~(LUT), 2443200 flip-flops~(FF), 2584 18Kb blocks of
block RAM~(BRAM), 2160 digital signal processing~(DSP) elements, and 24 MMCMs.

The accelerators instantiated in the SoC were obtained through
HLS of applications from the CHStone benchmark suite~\cite{Hara_2008ISCAS}.
We employed Cadence Xcelium 20.09 to simulate SoC instances,
AMD Vivado 2019.2 for HLS, RTL synthesis and implementation, and bitstream generation, and
Siemens proFPGA Builder 2019A-SP2 for FPGA programming.
Area and timing results refer to the post-implementation netlists of the prototypes,
while system-level statistics from the SoC execution are collected from the prototyped SoCs
through the run-time monitoring infrastructure described in Section~\ref{sec:methodology}.

\begin{figure}[t]
	\centering
	\includegraphics[width=\columnwidth]{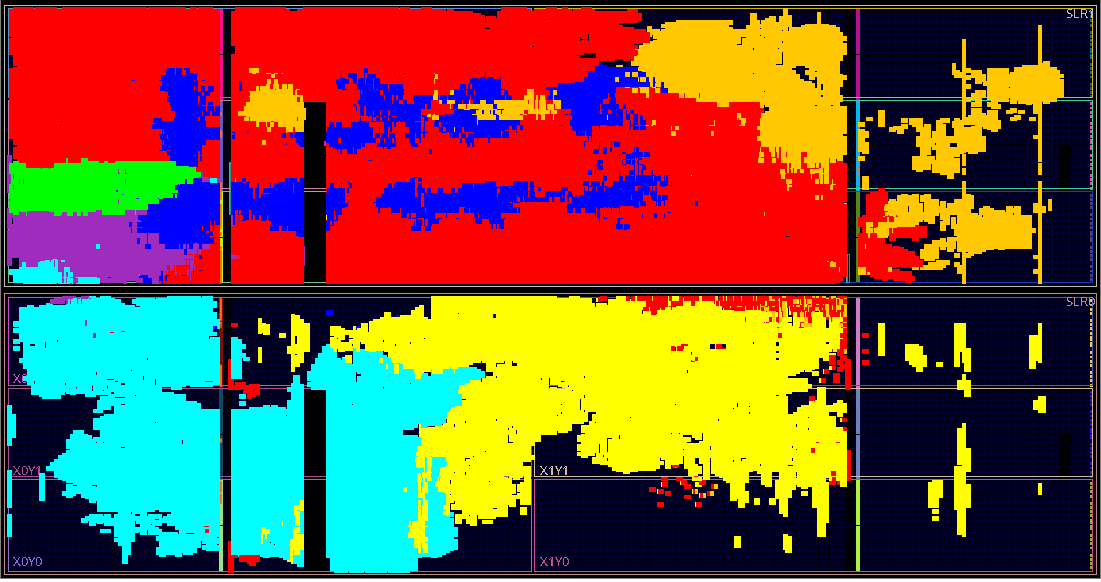}
	\caption{Floorplan of an instance of the Vespa SoC architecture.
		Legend:
		NoC in \textit{\textcolor{blue}{blue}},
		I/O in \textit{\textcolor{violet}{violet}},
		CPU in \textit{\textcolor{cyan}{cyan}},
		TGs  in \textit{\textcolor{red}{red}},
		MEM in \textit{\textcolor{green}{green}},
		A1~(\textit{dfsin}) in \textit{\textcolor{yellow}{yellow}},
		A2~(\textit{gsm}) in \textit{\textcolor{orange}{orange}}.
	}
	\label{fig:exp_floorplan}
\end{figure}

Without loss of generality,
we evaluated our methodology on 4-by-4 tile-based SoC instances
that feature a CVA6~\cite{Zaruba_2019TVLSI} CPU tile, a DDR memory one, and an auxiliary tile,
while the thirteen remaining tiles instantiate HLS-generated CHStone accelerators.
Eleven \textit{TG} tiles
generate traffic in the NoC interconnect and implement
\textit{dfadd} accelerators, which were
empirically observed to be memory-bound,
and two more accelerator tiles are placed in the \textit{A1} and \textit{A2} positions,
that are respectively close to~(\textit{A1}) and far from~(\textit{A2}) memory~(\textit{MEM}).
The SoC is split into five separate frequency islands,
namely,
the \textit{A1} and \textit{A2} accelerators,
the NoC interconnect and memory controller~(\textit{MEM}),
the traffic generation~(\textit{TG}) cores,
the CPU core~(\textit{CPU}), and
the auxiliary tile~(\textit{I/O}).
The DFS actuator of the NoC island implements a range of
operating frequencies comprised between 10MHz and 100MHz, while
the other four clock domains support clock signals with frequencies from 10MHz to 50MHz.
The clock frequency ranges include values at 5MHz steps. 
\figurename~\ref{fig:exp_floorplan} depicts the FPGA floorplan of an SoC instance.

\subsection{Multi-replica accelerator tiles: area-performance analysis}
\label{ssec:exp_accels}
Each baseline HLS-generated accelerator
occupies up to 1.4\%, 0.6\%, 1.0\%, and 3.8\% of the LUT, FF, BRAM, and DSP resources
available on the target FPGA, as listed in Table~\ref{tab:chstone_area},
thus justifying the possibility to improve their performance by
exploiting the multi-replica accelerator-tile architecture.
We explore therefore the instantiation of 2\texttimes-
and 4\texttimes-replication tiles and compare their
resource utilization and throughput.
The throughput is measured on the accelerator instantiated in the \textit{A1} tile,
with clock frequencies of 100MHz and 50MHz for the \textit{NoC}-\textit{MEM} and \textit{A1}-tile
islands, respectively, and with all \textit{TG} tiles disabled,
thus providing the best performance achievable by
the accelerator due to its proximity to memory and
no concurrent requests from the other computing tiles in the SoC.

The increase of LUT, FF, and BRAM utilization is shown to be quite smaller than
the replication factor, since such resources are
also being employed in the additional logic of the multi-replica accelerator tile.
On the contrary, DSP resources end up being around 2 and 4 times
as much as in the baseline versions.
Performance also scales correspondingly with
the replication of accelerators, as shown by the throughput columns in \tablename~\ref{tab:chstone_area}.
The 2\texttimes- and 4\texttimes-replication tiles have an average throughput increase
of 1.92\texttimes\, and 3.58\texttimes, respectively.
The larger average throughput improvement compared to the corresponding
increase in LUT, FF, and BRAM utilization justifies
our multi-replica accelerator-tile solution.

\begin{figure}[t]
	\centering
	\includegraphics[width=0.95\columnwidth]{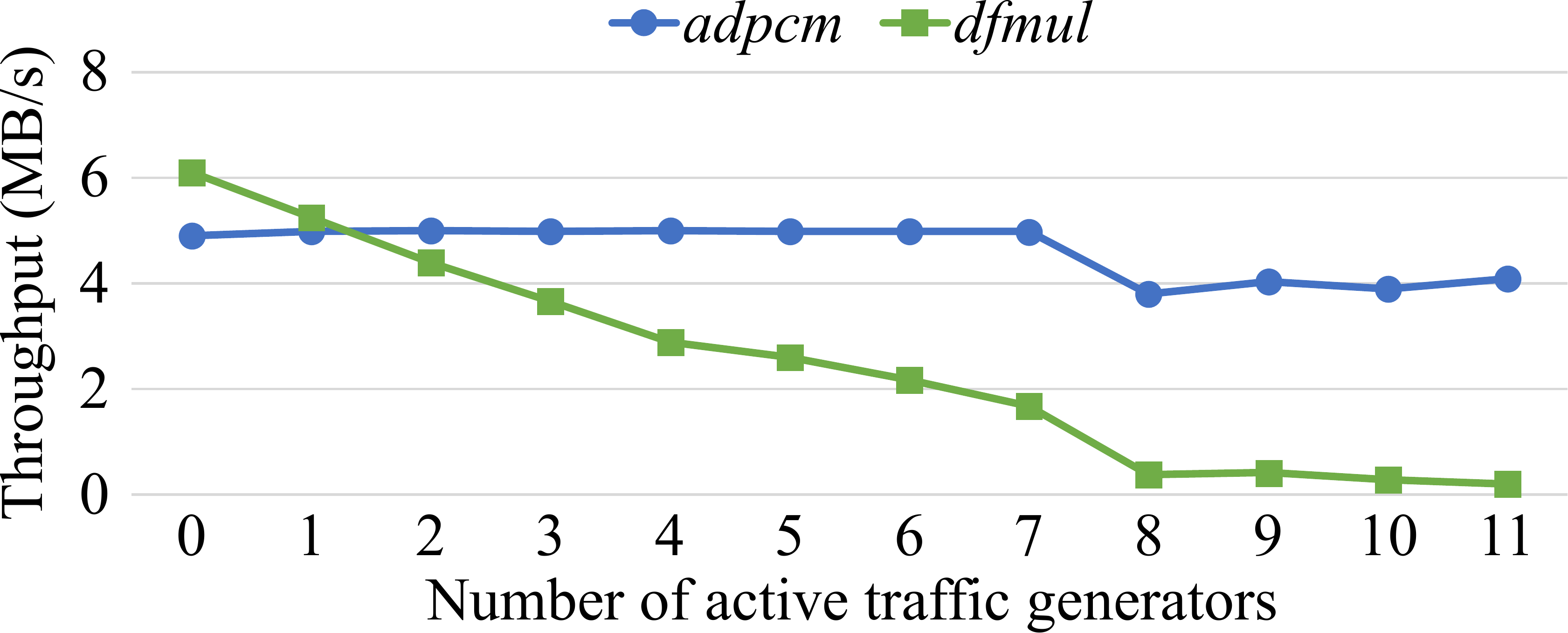}
	\caption{Throughput of 4\texttimes -replication
		compute-bound (\textit{adpcm}) and memory-bound (\textit{dfmul}) accelerators
		placed in the \textit{A2} tile at
		different numbers of active \textit{TG} cores.}
	\label{fig:rtt}
\end{figure}

\subsection{SoC accelerators: performance analysis}
\label{ssec:exp_soc}
We select \textit{adpcm} and \textit{dfmul} as representative examples of
compute- and memory-bound accelerators, respectively,
and analyze the relation between their throughput and
the amount of traffic in the NoC interconnect generated by \textit{TG} cores.
\figurename~\ref{fig:rtt} depicts the throughput of
the \textit{adpcm} and \textit{dfmul} accelerators
with a number of active \textit{TG} cores ranging between
0, i.e., when all of them are disabled, and
11, i.e., when all the \textit{TG} cores in the SoC are enabled.
The NoC interconnect runs at 10MHz while
the accelerators and \textit{TG} cores operate at 50MHz.

The compute-bound \textit{adpcm}
shows an almost constant throughput between 0 and 7 active \textit{TG} cores while,
on the contrary, in the same range of X axis values,
the throughput of the \textit{dfmul} accelerator drastically decreases,
thus highlighting the memory-bound nature of the latter.

\subsection{Memory: traffic analysis}
\label{ssec:exp_mem}
\begin{figure}[t]
	\centering
	\begin{subfigure}[t]{0.95\columnwidth}
		\includegraphics[width=\columnwidth]{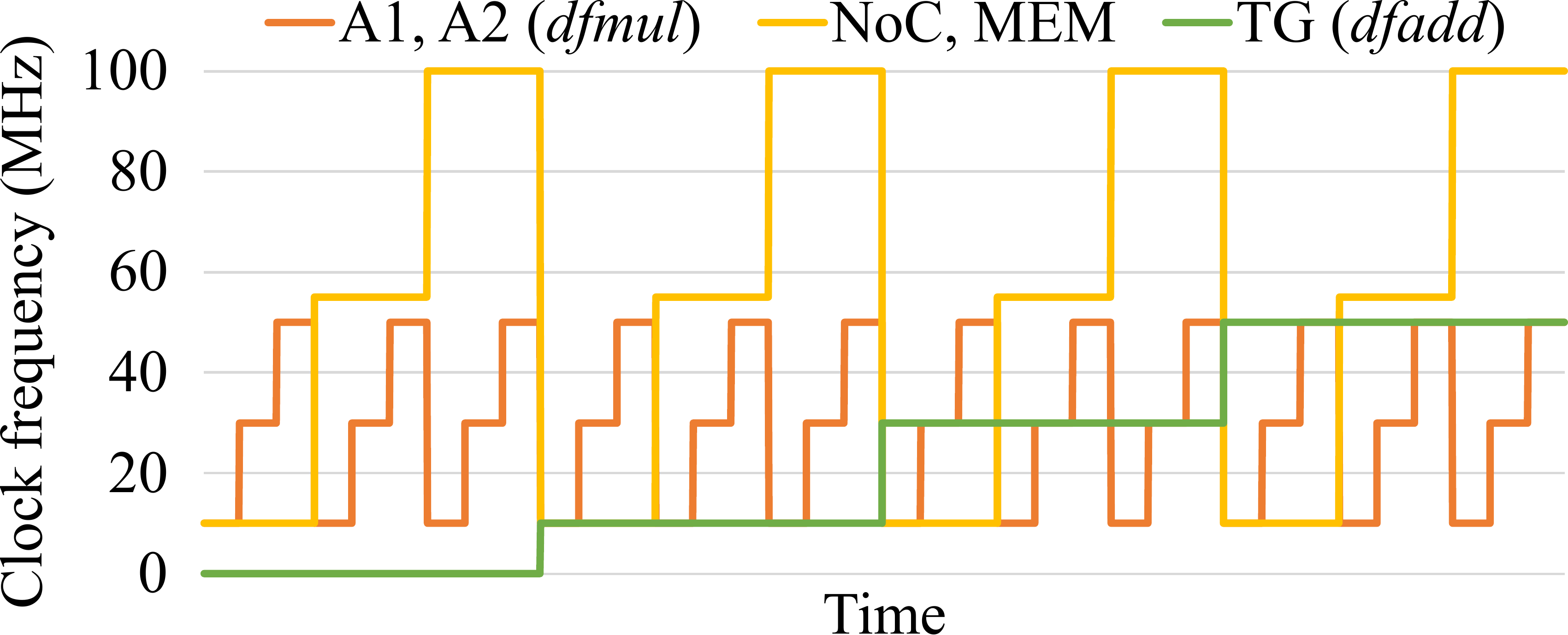}
		\caption{Clock frequency profiles.}
		\label{sfig:traffic_top}
		\vspace{0.35cm}
	\end{subfigure}
	\begin{subfigure}[t]{0.95\columnwidth}
		\includegraphics[width=\columnwidth]{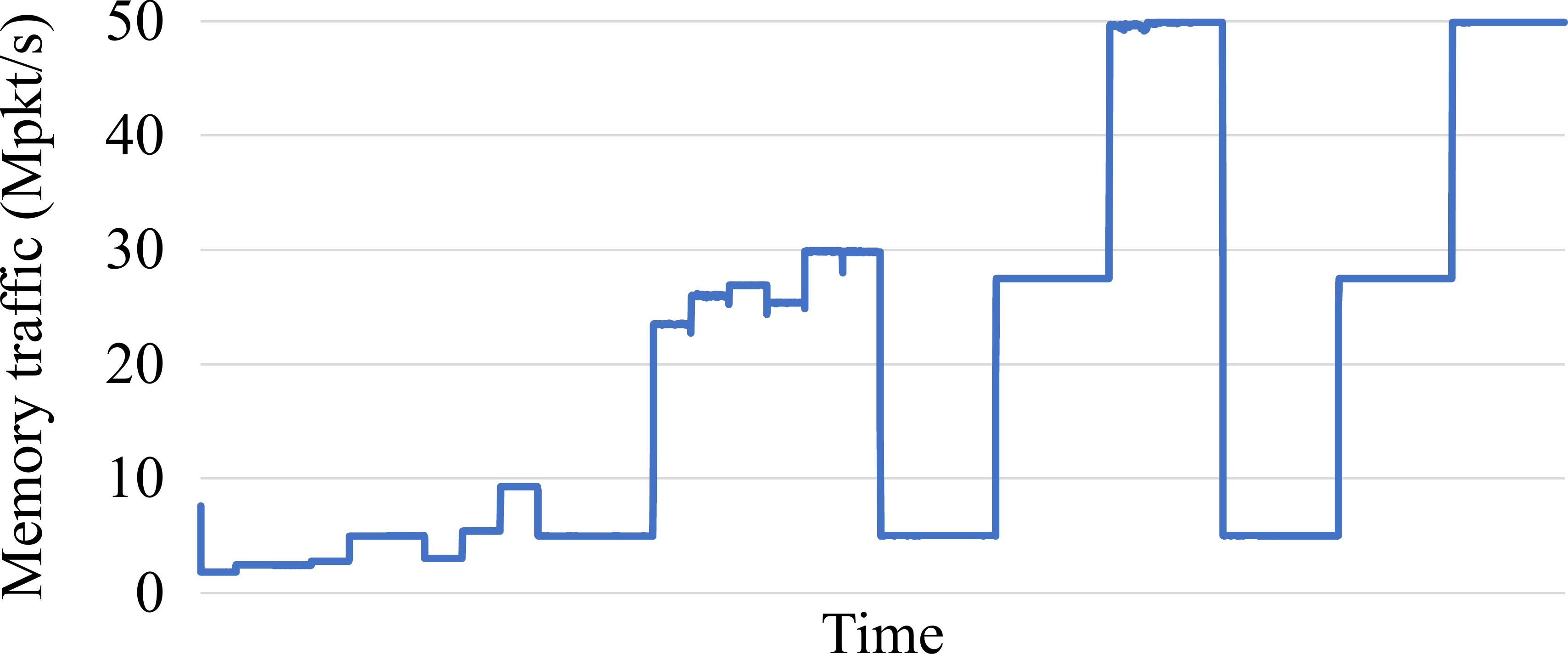}
		\caption{Memory incoming traffic.}
		\label{sfig:traffic_bottom}
	\end{subfigure}
	\caption{Memory incoming traffic while varying at run time
		the clock frequencies of the islands including
		the \textit{A1} and \textit{A2} tiles,
		the NoC interconnect and \textit{MEM} tile,
		and the \textit{TG} tiles.}
	\label{fig:traffic_profile}
\end{figure}

Finally, we show the possibility of evaluating the memory traffic depending on
the clock frequencies of the various frequency islands.
We consider a specific SoC instance with \textit{A1} and \textit{A2} tiles
both instantiating the memory-bound \textit{dfmul} and running concurrently.
\figurename~\ref{fig:traffic_profile} depicts the temporal evolution of incoming data packets to memory,
expressed as millions of packets per second~(Mpkt/s) in \figurename~\ref{sfig:traffic_bottom},
while varying the frequency islands' clock frequencies,
depicted in \figurename~\ref{sfig:traffic_top} in
red for the \textit{A1} and \textit{A2} tiles,
yellow for the NoC interconnect and memory controller, and
green for the \textit{TG} cores.

Varying the clock frequency of the \textit{A1} and \textit{A2} tiles between
10MHz, 30MHz, and 50MHz values is shown to have a negligible impact on the memory incoming traffic,
while increasing the operating frequency of the \textit{TG} cores drastically increases
the pressure on memory when the NoC interconnect and memory controller are
also running at a high clock frequency.
Such results demonstrate the possibility to analyze
the impact of DFS on the local and global traffic of the SoC.

%% file: 04_conclusions.tex
This paper introduced the open-source Vespa framework for designing large tile-based
multi-core heterogeneous SoCs. Vespa extends the open-source ESP
toolchain by
\begin{enumerate*}[label=\textit{\roman*)}]
	\item adding the possibility to configure the area-performance trade-off for each accelerator tile,
	\item partitioning the SoC into multiple frequency islands with independent true DFS actuators, and
	\item implementing a run-time monitoring infrastructure that enables the DSE of complex prototypes.
\end{enumerate*}

The experimental campaign, targeting a large AMD Virtex-7 2000 FPGA chip,
demonstrated its effectiveness in 
analyzing the area-performance trade-off by leveraging
the scalable throughput architecture, the DFS actuators, and
the run-time monitoring infrastructure, thus making
the proposed framework a valuable tool for designing large FPGA-based systems.

%% file: iccd.bbl
\begin{thebibliography}{1}
\providecommand{\url}[1]{#1}
\csname url@samestyle\endcsname
\providecommand{\newblock}{\relax}
\providecommand{\bibinfo}[2]{#2}
\providecommand{\BIBentrySTDinterwordspacing}{\spaceskip=0pt\relax}
\providecommand{\BIBentryALTinterwordstretchfactor}{4}
\providecommand{\BIBentryALTinterwordspacing}{\spaceskip=\fontdimen2\font plus
\BIBentryALTinterwordstretchfactor\fontdimen3\font minus
  \fontdimen4\font\relax}
\providecommand{\BIBforeignlanguage}[2]{{%
\expandafter\ifx\csname l@#1\endcsname\relax
\typeout{** WARNING: IEEEtran.bst: No hyphenation pattern has been}%
\typeout{** loaded for the language `#1'. Using the pattern for}%
\typeout{** the default language instead.}%
\else
\language=\csname l@#1\endcsname
\fi
#2}}
\providecommand{\BIBdecl}{\relax}
\BIBdecl

\bibitem{Galimberti_2022DSD}
A.~Galimberti, D.~Galli, G.~Montanaro, W.~Fornaciari, and D.~Zoni, ``Fpga
  implementation of bike for quantum-resistant tls,'' in \emph{2022 25th
  Euromicro Conference on Digital System Design (DSD)}, 2022, pp. 539--547.

\bibitem{Chen_2016JSSC}
Y.-H. Chen, T.~Krishna, J.~S. Emer, and V.~Sze, ``Eyeriss: An energy-efficient
  reconfigurable accelerator for deep convolutional neural networks,''
  \emph{IEEE Journal of Solid-State Circuits}, vol.~52, no.~1, pp. 127--138,
  2017.

\bibitem{Galimberti_2023ICECS}
A.~Galimberti, G.~Montanaro, and D.~Zoni, ``Hls-based acceleration of the bike
  post-quantum kem on embedded-class heterogeneous socs,'' in \emph{2023 30th
  IEEE International Conference on Electronics, Circuits and Systems (ICECS)},
  2023, pp. 1--4.

\bibitem{mantovani2020agile}
P.~Mantovani, D.~Giri, G.~Di~Guglielmo, L.~Piccolboni, J.~Zuckerman, E.~G.
  Cota, M.~Petracca, C.~Pilato, and L.~P. Carloni, ``Agile soc development with
  open esp,'' in \emph{Proceedings of the 39th International Conference on
  Computer-Aided Design}, 2020, pp. 1--9.

\bibitem{amid2020chipyard}
A.~Amid, D.~Biancolin, A.~Gonzalez, D.~Grubb, S.~Karandikar, H.~Liew,
  A.~Magyar, H.~Mao, A.~Ou, N.~Pemberton \emph{et~al.}, ``Chipyard: Integrated
  design, simulation, and implementation framework for custom socs,''
  \emph{IEEE Micro}, vol.~40, no.~4, pp. 10--21, 2020.

\bibitem{openpiton_2020}
\BIBentryALTinterwordspacing
J.~Balkind, T.-J. Chang, P.~J. Jackson, G.~Tziantzioulis, A.~Li, F.~Gao,
  A.~Lavrov, G.~Chirkov, J.~Tu, M.~Shahrad, and D.~Wentzlaff, ``Openpiton at 5:
  A nexus for open and agile hardware design,'' \emph{IEEE Micro}, vol.~40,
  no.~4, p. 22–31, jul 2020. [Online]. Available:
  \url{https://doi.org/10.1109/MM.2020.2997706}
\BIBentrySTDinterwordspacing

\bibitem{Zoni_2023CSUR}
\BIBentryALTinterwordspacing
D.~Zoni, A.~Galimberti, and W.~Fornaciari, ``A survey on run-time power
  monitors at the edge,'' \emph{ACM Comput. Surv.}, vol.~55, no. 14s, jul 2023.
  [Online]. Available: \url{https://doi.org/10.1145/3593044}
\BIBentrySTDinterwordspacing

\bibitem{Hara_2008ISCAS}
Y.~Hara, H.~Tomiyama, S.~Honda, H.~Takada, and K.~Ishii, ``Chstone: A benchmark
  program suite for practical c-based high-level synthesis,'' in \emph{2008
  IEEE International Symposium on Circuits and Systems (ISCAS)}, 2008, pp.
  1192--1195.

\bibitem{Zaruba_2019TVLSI}
F.~{Zaruba} and L.~{Benini}, ``The cost of application-class processing: Energy
  and performance analysis of a linux-ready 1.7-ghz 64-bit risc-v core in 22-nm
  fdsoi technology,'' \emph{IEEE Transactions on Very Large Scale Integration
  (VLSI) Systems}, vol.~27, no.~11, pp. 2629--2640, Nov 2019.

\end{thebibliography}
